# How balance and sample size impact bias in the estimation of causal treatment effects: A simulation study

*Andreas Markoulidakis*[*,1,3], *Peter Holmans*[4], *Philip Pallmann*[3], *Monica Busse*[3], *Beth Ann Griffin*[2]

[1]School of Medicine, Cardiff University, Cardiff, UK, [2]RAND Corporation, Arlington, VA, USA, [3]Centre for Trials Research, Cardiff University, Cardiff, UK, [4]Division of Psychological Medicine and Clinical Neurosciences, School of Medicine, Cardiff University, Cardiff, UK

**Correspondence:** *Corresponding author Andreas Markoulidakis, Email: MarkoulidakisA@cardiff.ac.uk

**Abstract**: Observational studies are often used to understand relationships between exposures and outcomes. They do not, however, allow conclusions about causal relationships to be drawn unless statistical techniques are used to account for the imbalance of con- founders across exposure groups. Propensity score and balance weighting (PSBW) are useful techniques that aim to reduce the imbalances between exposure groups by weighting the groups to look alike on the observed confounders. Despite the plethora of available methods to estimate PSBW, there is little guidance on what one defines as adequate balance, and unbiased and robust estimation of the causal treatment effect is not guaranteed unless several conditions hold. Accurate inference requires that 1) the treatment allocation mechanism is known, 2) the relationship between the base- line covariates and the outcome is known, 3) adequate balance of baseline covariates is achieved post-weighting, 4) a proper set of covariates to control for confounding bias is known, and 5) a large enough sample size is available. In this article we use simulated data of various sizes to investigate the influence of these five factors on statistical inference. Our findings provide evidence that the maximum Kolmogorov- Smirnov statistic is the proper statistical measure to assess balance on the baseline covariates, in contrast to the mean standardised mean difference used in many appli- cations, and 0.1 is a suitable threshold to consider as acceptable balance. Finally, we recommend that $60 − 80$ observations, per confounder per treatment group, are required to obtain a reliable and unbiased estimation of the causal treatment effect.

# Introduction

Randomized controlled trials (RCTs) are the gold standard in the estimation of causal treatment effects in clinical trials, because randomization of a large enough sample usually creates two groups that are well-balanced on both observed and unobserved pre-treatment confounders. Observational studies on the other hand compare two or more groups (e.g., *treatments* and *control*) where the allocation mechanism of individuals to groups is unknown, and typically not random. These are employed in a range of circumstances, such as where an RCT design is not feasible or where there are routine datasets or cohorts providing valuable information about outcomes in relation to certain exposures. Group assignment might be due to factors that the researcher cannot control, such as underlying conditions of the individual. These differences between the groups could induce bias in terms of estimated effects of treatment. For example, consider a study that aims to estimate the causal effect of physical activity on disease progression among a sample of individuals living with a particular disease using observational study data.The likelihood that an individual engages in regular physical activity at study entry might in part be influenced by their disease severity (e.g., those with more severe disease may struggle with motivation and/or ability to engage in regular physical activity) and any subsequent comparison between those who did and did not exercise will be distorted, as it will compare groups with different pre-treatment levels of disease severity, potentially overstating the benefits of exercise for individuals living with the disease.

The estimation of accurate causal treatment effects (Holland 1986) in observational studies is heavily discussed in the literature, with propensity score (PS) (Rubin 1977) techniques being extensively used to reduce the confounding bias. The PS (Rubin 1977) is the probability of an individual's allocation to the treatment group, given their observed baseline (pre-treatment) characteristics. The PS can be used to create pseudo-randomized comparable treatment groups by either weighting, matching, adjusting, or stratifying on the PS. PS methods reduce the bias in the estimation of the causal treatment effect due to the observed confounders by minimizing the imbalance of known and observed confounders between the treatment groups.

We focus our current study on PS and balancing weighting, though the conclusions are likely to generalize to other uses of the PS and balancing weights. Despite the wealth of investigation into PS weighting (Elze et al. 2017; Olmos and Govindasamy 2015; Posner and Ash 2012; Harder, Stuart, and Anthony 2010), there are still a number of uncertainties that limit best practice recommendations. Accurate inference based on observational data is conditional on several assumptions (e.g., positivity and unconfoundedness) which, if met, guarantee an unbiased and robust estimation of the causal treatment effect. The vast majority of methods (Elze et al. 2017; Olmos and Govindasamy 2015; Posner and Ash 2012; Harder, Stuart, and Anthony 2010; Pirracchio, Resche-Rigon, and Chevret 2012) assume that 1) the treatment allocation mechanism is known, 2) the relationship between the baseline covariates and the outcome is known, 3) adequate balance of baseline covariates is achieved post-weighting, 4) a proper set of covariates to control for confounding bias is known, and 5) a large enough sample size is available. In this article, we use simulated data, inspired by two previous studies (Setoguchi et al. 2008; Setodji et al. 2017), to examine the extent compliance with these five conditions is necessary for robust causal effect estimation.

In terms of exploring the impact of sample size on the accuracy of the estimation of the causal treatment effect, we note that there has been a rich body of literature using the same simulation models as used here to investigate the performance of different PS and balancing weights estimation algorithms (Setoguchi et al. 2008; Setodji et al. 2017; Choi, Shing Wan and Mak, Timothy Shin-Heng and O'Reilly 2020; J. Lee and Little 2017; Abdia et al. 2017; Wyss et al. 2014; Pirracchio, Petersen, and van der Laan 2015; Gharibzadeh et al. 2018; Xie et al. 2019). In each case, the studies typically only used two different sample sizes ($n = 2000$ and $n = 10000$ observations in total). Related work (Gharibzadeh et al. 2018; Wyss et al. 2014; Xie et al. 2019; Harvey et al. 2017) investigating the relative strength of different PS and balancing weights estimation algorithms also focused on the performance of the algorithms for large samples, giving little attention to issues arising in the analysis of small samples. Based on these studies, a sample size of $n = 2000$ appears sufficient to effectively estimate the causal treatment effect, even when the true outcome model deviates significantly from the one used in the statistical analysis. There is little discussion in the literature about the potential performance of different PS and balancing weight methods when applied to smaller samples (Pirracchio, Resche-Rigon, and Chevret 2012), which are usual when studying rare diseases or when there are limited means to collect data. Here, we simulated samples of several sizes (from $n = 40$ to 2000), in order to determine the minimum sample size needed per confounder and treatment group that allows for weighting methods to achieve a reasonable bias in the estimated causal treatment effect. In the occasions that small sample sizes are considered (Pirracchio, Resche-Rigon, and Chevret 2012; Li and Li 2021), the true propensity score and outcome models (the true underlying relationship between the baseline covariates and the treatment/outcome) are simple logistic regression and/or linear regression with the main effects only. In such cases, there is evidence that even a regression model with no control for confounding bias could yield an estimation of the causal treatment effect with low relative bias (Setoguchi et al. 2008; Setodji et al. 2017). In this paper, we consider more complicated treatment allocation and outcome models, in a way that the logistic and linear regression models — that are traditionally used for PS and outcome estimation, respectively — do not reflect the truth. Our inspiration for exploring such small sample sizes is the limitations that often arise in real-world data, especially when they are related to rare diseases (Day et al. 2018; Mitani and Haneuse 2020) and there is a large number of observed confounding variables.

Even though PS and balancing weights have been used for decades (Setoguchi et al. 2008; Setodji et al. 2017; Griffin et al. 2017), the definition of adequate balance is not yet well established across the community. The most common measure of balance used in the literature is the standardized mean difference (SMD) between treatment and control groups for each of the observed pre-treatment confounders used in the analysis (Griffin et al. 2017; Franklin et al. 2014; Zhang et al. 2019; Austin 2009; Stuart, Lee, and Leacy 2013; Griffin et al. 2014). In this study, we investigate the strengths and limitations of using the SMD versus the Kolmogorov-Smirnov (KS) statistic (Gail and Green 1976), which is rarely used in studies. We also investigate optimal thresholds to minimise bias in treatment effect estimation for both statistics. Typically, a 0.2 (Abdia et al. 2017) balance threshold for the SMD is suggested as sufficient to guarantee unbiased estimation of the causal treatment effect, but our work shows that stricter rules should apply (Setoguchi et al. 2008; Setodji et al. 2017; Griffin et al. 2017),.

Selecting which covariates to control for in the PS and balancing weights is an issue that has been well discussed previously, with most authors arguing in favor of controlling for the true confounders as well as covariates causally linked only to the outcome since the inclusion of covariates related only to the treatment allocation could inflate the bias of the estimate and the mean squared error (MSE) (Brookhart et al. 2006; Hirano and Imbens 2001; Adelson et al. 2017; Perkins et al. 2000). We also study the role of variable selection in our simulations with more restricted sample sizes by evaluating the impact of every possible combination of confounders on the outcome estimation, making suggestions regarding which sets of covariates should be prioritized for controlling.

The remainder of the article is organised as follows. *Section* 2 provides a brief overview of the basic principles for estimation of the causal treatment effect. In *section* 3 we briefly review the algorithms that we use to obtain PS and balancing weights and the statistics we use to evaluate the balance. *Section* 4 discusses the choice of baseline covariates used in the estimation of the PS and balancing weights. In *section* 5 we define the different sets of confounders used in the simulations and present the data simulation framework. We report and discuss our findings in *Section* 6.

Additionally, it is usually easier to achieve an acceptable level of balance on SMD rather than KS statistic, since the former concerns only the mean value of the groups (one-dimensional measure), while the latter concerns the entire distribution of the two groups.

Our study considers different sets of covariates to control for confounding bias, thus nicely revealing the caveats that occur when *wrong* covariates are chosen to control for confounding bias and also how misleading sometimes the balance measures could be, if covariates not related to the outcome are deployed. All these set-ups are evaluated across data-sets of different size, to provide a *rule-of-thumb* regarding the number of participants one needs for each confounder to achieve adequate balance.

## 2. Overview of Causal Modeling

In the causal inference framework, we are often interested in evaluating the performance of a new treatment compared to one traditionally used or an untreated situation, which is referred to as reference or control treatment. This measure is usually called the *average treatment effect* (Holland 1986). In this article, we focus on studies with two groups, named control and treatment.

In Rubin's causal model (RCM), a formal mathematical framework for causal inference (Rubin 1978; 1976; 1979; 1977; Rosenbaum and Rubin 1983), every individual $i = 1, \ldots, n$ has a potential outcome conditional on the presence of treatment ($\{Y_{i|T=1}\}$), and a potential outcome conditional on the absence of treatment ($\{Y_{i|T=0}\}$). The treatment effect for each individual is defined as the difference between the two potential outcome effects

$$\tau_i = Y_{i|T=1} - Y_{i|T=0}.$$

It is impossible to observe both outcomes for the same individual since every individual is exclusively assigned either to the control or treatment group. Thus, we are able to observe the outcome in the presence of treatment $Y_{i|T=1}$ only for individuals in the treatment group ($Y_{i|T=1}|T_i = 1$) and the outcome in the absence of treatment $Y_{i|T=0}$ only for individuals in the control group ($Y_{i|T=0}|T_i = 0$).

In this article, we estimate the average treatment effect on the treated population (ATT) (Rosenbaum and Rubin 1983), which is defined as

$$ATT = E[Y_{i|T=1}|T = 1] - E[Y_{i|T=0}|T = 1],$$

where $E[Y_{i|T=1}|T = 1]$ is the potential outcome under treatment, given that the individual receives treatment, while $E[Y_{i|T=0}|T = 1]$ is the potential outcome under control, given that the individual receives treatment. The latter quantity is not possible to be observed — it is actually the outcome of an individual of the control group, had it been assigned to the treatment group. Since it is impossible to observe this quantity, balancing weights are used to adjust the observed outcomes of the control group, supposed they received treatment, to match the baseline characteristics of the treatment group.

ATT measures the effect of the treatment only among individuals similar to those in the treatment group. Other quantities of interest are the average treatment effect on the entire population (ATE) and the average treatment effect on the control population (ATC), which express the average causal treatment effect for the entire population of individuals in both the treatment and control groups and the effect of the treatment among only individuals similar to those in the control group, respectively.

In order to obtain accurate estimates of $E[Y_{i|T=1}|T=0]$ and $E[Y_{i|T=0}|T=1]$, RCM requires two assumptions to hold, named *strong ignorability* and *stable unit treatment value assumption* (SUTVA) (Rosenbaum and Rubin 1983).

*Strong ignorability* assumes that the treatment assignment mechanism is independent of the distributions of the outcome, given the observed covariates $X$ on the baseline(($Y_{i|T=0}, Y_{i|T=1} \perp T|X$ — unconfoundedness). This assumption also requires that every individual, given the values of the covariates $X$ on the baseline, has a positive probability greater than 0 and less than 1 to be assigned either on treatment or control group ($0 < P(T_i = 1) < 1$ for every $i$). In other terms, the individuals in the two treatment groups should overlap on the baseline characteristics, thus there are representatives from both groups for each value of baseline covariates. If the control and intervention group are properly balanced after the weighting, this is a sufficient indication that strong ignorability has been achieved on the treatment assignment given the observed covariates.

SUTVA states that the outcome value $Y_{i|T=k}$ corresponding to individual $i$ for treatment $k$ ($k \in \{0,1\}$) is unique. This assumption implies that the distribution of potential outcomes for each individual is independent of the potential outcome of another individual. Additionally, it implies that all possible values of treatment status are represented (Cox and Cox 1958).

*Inverse probability weighting* (IPW) (Robins, Hernan, and Brumback 2000) assigns a weight to each individual, and then the treatment effect is computed from the weighted mean effect on control and treatment group, as follows:

$$\widehat{E[Y_0]} = \frac{\sum_{i=1}^{n_0} w_i^0 Y_{i|T=0}}{\sum_{i=1}^{n_0} w_i^0}, \qquad \widehat{E[Y_1]} = \frac{\sum_{i=1}^{n_1} w_i^1 Y_{i|T=1}}{\sum_{i=1}^{n_1} w_i^1}.$$

These weights are often derived from an individual's PS (Imai and Ratkovic 2014; Ridgeway et al. 2017; Olmos and Govindasamy 2015), however, recent algorithms allow direct estimation of the weights (Hainmueller 2012), subject to a set of restrictions, to achieve better balance among the treatment groups — see section 3.1.

It is very common practice to use a multivariable regression to adjust for the weights, which ideally includes all of the observed confounders used in the estimation of the weights (Ridgeway et al. 2017; Austin 2011) — and maybe other covariates that were not used to control for confounding bias. When using PS weights, this approach is called a *doubly robust estimator* of the causal treatment effect (Bang and Robins 2005; Kang, Schafer, and others 2007; Chattopadhyay, Hase, and Zubizarreta 2020; Zhao and Percival 2016). The estimated treatment effect is consistent so long as either the PS weight model or the multivariable outcome model is correctly specified.

## 3. Weighting & Balance Evaluation

### Propensity Score & Balancing Weights

We will use four main algorithms to compute PS and balancing weights, named Logistic Regression (LR) (Agresti 2018; Wright 1995), Covariate Balance Propensity Score (CBPS) (Imai and Ratkovic 2014), Generalized Boosted Model (GBM) (McCaffrey, Ridgeway, and Morral 2004) and Entropy Balance (EB) (Hainmueller 2012).

### *Logistic Regression (LR)*

The simplest and most commonly used parametric method to estimate the PS of each individual is LR (Agresti 2018) since treatment assignments are often binary. The LR model for estimating the PS assumes that the *logit* of the probability of receiving treatment is a linear combination of the covariates [1]. The main caveat with the LR is that the PS model can often be misspecified, leading to biased estimates of treatment effects. More complex relationships of the treatment with the baseline covariates could be considered (including higher orders and/or interactions), but this requires the user to be familiar with parametric models to define them manually.

### *Covariate Balance Propensity Score (CBPS)*

CBPS (Imai and Ratkovic 2014) is a parametric method which is deployed to overcome some of the potential misspecifications of the LR. CBPS still assumes the same relationship between the treatment status and the baseline covariates, like the LR, however, it adds further constraints to achieve a good balance between the treatment groups. The set of constraints considered in the basic version of CBPS targets to balance the first moment of the distributions of the baseline covariates of the two groups (the means). As a consequence, this method is robust to mild model misspecification about balancing confounders compared to standard LR (Imai and Ratkovic 2014; Setodji et al. 2017; Choi et al. 2019; Wyss et al. 2014; Xie et al. 2019).

Further extensions of CBPS can achieve balance of higher moments (Huang MY Vegetabile B, n.d.; Markoulidakis et al. 2021). It is possible to impose restrictions that require the first $m$ moments to be matched, by replacing the original covariates with orthogonal polynomials of degree $m$. This transformation is possible only on continuous covariates, and it further increases the number of confounders that the PS model should control for — e.g. if we have five continuous confounders, setting $m = 3$ would increase the number of confounders from five to $3 \cdot 5 = 15$.

---

[1] An extension of LR is Multinomial Logistic Regression which could be used to estimate generalized PS if there are more than two treatment conditions.

In this study, we consider three versions of the CBPS algorithm, controlling for $m = 1,2$ and $3$ moments (noted as $CBPS\#1$, $CBPS\#2$ and $CBPS\#3$, respectively).

*Generalized Boosted Model (GBM)*

GBM is a non-parametric machine learning approach to estimating PS weights. It predicts the binary treatment indicator by fitting a piecewise-constant model, constructed as a combination of simple regression trees (Burgette, McCaffrey, and Griffin in press, (Burgette, McCaffrey, and Griffin 2015; Ridgeway 1999; Ridgeway et al. 2017)), namely *Recursive Partitioning Algorithms* and *Boosting*. To develop the PS model, GBM uses an iterative, *forward stagewise additive algorithm*. Starting with the PS equal to the average of treatment assignment on the sample, the algorithm starts by fitting a simple regression tree to the data to predict treatment from the covariates by maximizing the following function

$$l(x) = \sum_{i=1}^{N} T_i\, g(X_i) - log(1 + exp(g(X_i))),$$

where $g(X_i)$ is the $logit$ of treatment assignment. In each iteration the algorithm splits the nodes of the tree with respect to the criterion that one wishes to minimize (Ridgeway et al. 2017) (this is usually the *mean SMD* or the *maximum KS* value). The algorithm will stop either when adequate balance has been achieved, or the maximum number of iteration has been reached.

*Entropy Balance (EB)*

EB (Hainmueller 2012) is a method that estimates the weights directly rather than the PS of the individuals. The method attempts to achieve exact balance (difference of moments across the treatment groups equal to 0) on as many moments as defined by the user. EB calculates weights through a re-weighting scheme until the adequate balance in the pre-selected moments is achieved, attempting to match the first $m$ moments of the distributions of the two groups.

In this study, we consider three versions of the EB algorithm, controlling for $m = 1,2$ and $3$ moments (noted as $EB\#1$, $EB\#2$ and $EB\#3$, respectively).

## Balance Evaluation

Once we have an estimation of the PS and balancing weights, it is important to evaluate the balance on the two groups achieved. To do so we will use the *standardized mean difference* (SMD) and *Kolmogorov-Smirnov statistic* (KS).

The SMD (Austin 2009; Franklin et al. 2014) is a measure of the distance of the means of two groups. It is defined as the difference of the means, divided by an estimate of the standard deviation for a given covariate. We will concentrate on the *Absolute SMD*, whose values are non-negative, with lower values corresponding to better balance. Initially, 0.2 was recommended as a threshold (Abdia et al. 2017) to define groups as balanced. However, there is more recent evidence that a more conservative threshold (0.1) should be considered (Griffin et al. 2017; Austin 2009; Zhang et al. 2019; Griffin et al. 2014; Stuart, Lee, and Leacy 2013).

KS is a test statistic used in a procedure (Gail and Green 1976) which tests the hypothesis that two samples are from the same distribution. It takes values in $[0,1]$, and lower values indicate that the two distributions are more similar. In contrast to SMD, KS quantifies the similarity of the entire distribution of the two groups, rather than the means only. There is no clear guidance on what is the best threshold for the KS but values over 0.1 would be considered notably large and thus, we propose to use 0.1 as the threshold for balance for the KS as well as the SMD. It is also expected that once balance across distributions of the groups has been achieved (KS value under 0.1), balance across the means is also held (SMD value under 0.1).

Finally, we compute the *effective sample size* (Ridgeway et al. 2017) (ESS) for each algorithm. This is a measure of the sample power lost by weighting, and corresponds to the sample size of an unweighted analysis that would give the same power as the weighted analysis.

In this paper, we use Logistic Regression ($LR$), Generalized Boosted Model minimizing the mean SMD value ($GBM_{ES}$), Generalized Boosted Model minimizing the maximum KS value ($GBM_{KS}$), Covariate Balance Propensity Score controlling for moments $m = 1,2,3$ ($CBPS\#1$, $CBPS\#2$, $CBPS\#3$, respectively), and Entropy Balancing controlling for moments $m = 1,2,3$ ($EB\#1$, $EB\#2$, $EB\#3$, respectively). $CBPS\#2$, $CBPS\#3$, $EB\#2$ and $EB\#3$, attempt to control for higher moments than just the mean, so this could match the higher-order moments existing on the true relationship between baseline covariates and the treatment allocation. Since our aim is to provide general guidelines for balance, we will use all the algorithms used in (Markoulidakis et al. 2021) in our analyses. We do not intend to compare the performance of the algorithms, and thus we do not explicitly recommend one algorithm over the others. Instead, we encourage users to use more than one algorithm to evaluate balance on the baseline covariates and use the one that achieves the best trade-off between balance and ESS to model the outcome .

## 4.    Understanding the Importance of Variable Selection

The lack of random allocation mechanism among the treatment groups in observational studies (Holland 1986) incurs confounding bias, which PS and balancing weights are intended to minimize (Rosenbaum and Rubin 1983). A key consideration when deploying PS and balancing weights to control for confounding bias is the selection of variables that will be considered as confounders (predictors of both the outcome and the treatment status) .

The covariates that affect both the treatment allocation and the outcome value (Leite 2016) are the *true confounders*, and should always be included in the estimation of the PS and balancing weights (Leite 2016; Hirano and Imbens 2001; Brookhart et al. 2006; Perkins et al. 2000).

In the early stages of using PS and balancing weights to control for confounding bias, it was made explicit that all true confounders should be included in the treatment allocation model estimation (Brookhart et al., 2006; Hirano & Imbens, 2001; Perkins et al., 2000) — the model used for the estimation of PS/balancing weights. However, recent literature suggests that including covariates that are highly related to the treatment allocation but not to the outcome could increase the bias of the causal treatment effect estimation (Adelson et al. 2017). Also, covariates related to the outcome, but not to the treatment allocation should be included in the model, even though they would not increase the bias, but they could rather increase the precision of the estimated exposure effect (Brookhart et al. 2006; Patrick et al. 2011). Thus, it is suggested that it is preferable to control for covariates that are related to the outcome only, rather than related only to the treatment allocation (Pearl 2011) — of course, true confounders should always be prioritized for inclusion in the treatment allocation model.

Others suggest including in the PS model all the baseline covariates that are not balanced (Nguyen et al. 2017) — this is all the baseline covariates with an SMD over 0.1. This suggestion is dubious, since it is possible that covariates related to the treatment allocation but not the outcome have SMD over 0.1. Including such a covariate could increase the bias (Adelson et al. 2017), which is not a desirable outcome.

It is still unclear whether the covariates used in the treatment allocation model and the outcome model should be the same (Holland 1986; Rosenbaum and Rubin 1983). Even though in most applications this is the case — the same set of covariates is used in both models (Setodji et al. 2017; Setoguchi et al. 2008). Instead, Hirano and Imbens (Hirano and Imbens 2001), made explicit that the covariates of the two models are not necessarily the same. This is also implied by their proposed procedure to choose covariates for each regression model. They propose a step-wise method for variable selection, where covariates are added to and removed from each model until only statistically significant covariates are included in each model. Since the procedure is performed separately for the treatment allocation and the outcome model, and the choice of covariates is only based on their statistical significance as this is measured by the model, it is apparent that the covariates included in the two models are not necessarily the same.

## 5. Overview of Simulation

To investigate the features we are interested in, we used simulated data, based on the simulation study performed by Setoguchi and colleagues (Setoguchi et al. 2008). The analysis considers a binary treatment variable $T$ and ten pre-treatment covariates $(X_1, X_2, \ldots, X_{10})$.

To generate the ten pre-treatment covariates $(X_1, X_2, \ldots, X_{10})$, initially we generate six independent covariates following a standard normal distribution $(X_1, X_2, X_3, X_4, X_7, X_{10})$, with zero mean and unit variance. The remaining four covariates $(X_5, X_6, X_8, X_9)$ are generated using the correlation matrix:

$$\rho = \begin{bmatrix} 1 & 0 & 0 & 0 & 0.2 & 0 & 0 & 0 & 0 & 0 \\ 0 & 1 & 0 & 0 & 0 & 0.9 & 0 & 0 & 0 & 0 \\ 0 & 0 & 1 & 0 & 0 & 0 & 0 & 0 & 0 & 0 \\ 0 & 0 & 0 & 1 & 0 & 0 & 0 & 0.2 & 0 & 0 \\ 0.2 & 0 & 0 & 0 & 1 & 0 & 0 & 0 & 0.9 & 0 \\ 0 & 0.9 & 0 & 0 & 0 & 1 & 0 & 0 & 0 & 0 \\ 0 & 0 & 0 & 0 & 0 & 0 & 1 & 0 & 0 & 0 \\ 0 & 0 & 0.2 & 0 & 0 & 0 & 0 & 1 & 0 & 0 \\ 0 & 0 & 0 & 0.9 & 0 & 0 & 0 & 0 & 1 & 0 \\ 0 & 0 & 0 & 0 & 0 & 0 & 0 & 0 & 0 & 1 \end{bmatrix}$$

Finally, six covariates are converted to binary variables $(X_1, X_3, X_5, X_6, X_8, X_9)$ — using random numbers between 0 and 1 (see *Appendix* A).

## Treatment Allocation

The binary treatment variable, $T$ was modeled using *logistic regression* as a function of $X_i$. The formula used to compute the true propensity score ($P(T|X_i)$), is

$$P[T = 1|X] = \frac{1}{1 + e^{-(A+B+C)}}$$

$$\begin{aligned}
A &= \beta_0 + \beta_1 X_1 + \beta_2 X_2 + \beta_3 X_3 + \beta_4 X_4 + \beta_5 X_5 + \beta_6 X_6 + \beta_7 X_7, \\
B &= 0.5 \cdot \beta_1 X_1 X_3 + 0.7 \cdot \beta_2 X_2 X_4 + 0.5 \cdot \beta_3 X_3 X_5 + 0.7 \cdot \beta_4 X_4 X_6 + 0.5 \cdot \beta_5 X_5 X_7 \\
&\quad + 0.5 \cdot \beta_1 X_1 X_6 + 0.7 \cdot \beta_2 X_2 X_3 + 0.5 \cdot \beta_3 X_3 X_4 + 0.5 \cdot \beta_4 X_4 X_5 + 0.5 \cdot \beta_5 X_5 X_6, \\
C &= \beta_2 X_2 X_2 + \beta_4 X_4 X_4 + \beta_7 X_7 X_7,
\end{aligned}$$

corresponding to *scenario G* of Setoguchi (Setoguchi et al. 2008), with strong non-linearity and non-additivity. The true values of the coefficients used are:

$$(\beta_0, \beta_1, \beta_2, \beta_3, \beta_4, \beta_5, \beta_6, \beta_7) = (0, 0.8, -0.25, 0.6, -0.4, -0.8, -0.5, 0.7).$$

Setoguchi (Setoguchi et al. 2008) and others (Abdia et al. 2017; Setodji et al. 2017; Choi et al. 2019; Gharibzadeh et al. 2018; B. K. Lee, Lessler, and Stuart 2010; Pirracchio, Petersen, and van der Laan 2015; Wyss et al. 2014; Xie et al. 2019) typically used seven treatment models, each expressing a different strength of linearity and additivity, since the main goal of those articles was to explore the relative performance of several different PS and balancing weights estimation algorithms. Our goal is not to determine which algorithm performs best in terms of PS estimation, but to assess the overall quality of the results produced by the different candidate methods based on different levels of balance, sample size, and sets of confounders. Thus, we focus on only one treatment allocation mechanism — one that is non-linear and non-additive, in order to ensure our findings reflect potential modelling issues encountered in real world applications.

For the simple linear and additive treatment allocation model (this corresponds to *Scenario 1* of Setoguchi (Setoguchi et al. 2008; Setodji et al. 2017), every algorithm can provide a valid model for the estimation of PS and balancing weights. The doubly-robust method traditionally used for the estimation of the causal treatment effect assumes that at least one of the treatment allocation and outcome models is properly specified. In real-world study data, it is impossible to know the exact relationship between the confounders and the treatment/outcome covariate. Thus, one typically uses a simple linear (or logistic) regression model to express the relation of the outcome with the confounders when the outcome is continuous (binary, respectively). Modern algorithms provide a wider range of parametric (Olmos and Govindasamy 2015; Imai and Ratkovic 2014; Ridgeway et al. 2017; Hainmueller 2012) and non-parametric (Hainmueller 2012) functions to model the relationship between the confounders and the treatment, however, it is still not possible to test whether the assumptions of the treatment allocation model hold.

## Outcome Models

In this article, we consider four different models to generate the outcome. We do so to understand the sample size and balance level needed to obtain a good estimation of the causal treatment effect under a range of outcome models. In most applied analyses, researchers will tend to use additive, parametric models to estimate the causal effects of a binary treatment on outcomes (e.g., linear regression models to model continuous outcomes and logistic regression to model binary outcomes which control for main effects of the key confounders) (Robins, Hernan, and Brumback 2000; Hernán, Brumback, and Robins 2000; Setoguchi et al. 2008) and we want to understand how well such estimation approaches will perform as a function of the true underlying relationship between the outcome and the pre-treatment confounders. When one is working with real data, it is almost impossible to know the true relationship between the outcome and the predictors, thus these models are sensible candidates to help us quantify the impact the true underlying outcome model form may have on inferences.

Here, we use one binary model and three continuous ones to generate the true outcome, which deviates from the models used to estimate the treatment effect in the outcome analysis to varying degrees.

The binary model fits a *logistic regression* as a function of $X$ and $T$ as follows:

- **Outcome 1.**

$$P(Y = 1|T, X) = \frac{1}{1 + e^{-(\alpha_0 + \alpha_1 X_1 + \alpha_2 X_2 + \alpha_3 X_3 + \alpha_4 X_4 + \alpha_5 X_8 + \alpha_6 X_9 + \alpha_7 X_{10} + \gamma T)}}$$

where the continuous value of $P(Y = 1|T, X)$ is dichotomized — using random numbers between 0 and 1 (see *Appendix* A). This corresponds to the original outcome model proposed by Setoguchi (Setoguchi et al. 2008). *Outcome model* 1 is a truly additive model and thus our estimation of the treatment effect using simple main effects logistic regression should perform well so long as we include the correct set of pre-treatment covariates as independent variables in the regression model for $Y$.

The three continuous outcome models are

- **Outcome 2.**

$$Y = \alpha_0 + \gamma T + e^{\alpha_1 X_1 + \alpha_2 X_2 + \alpha_3 X_3} + \alpha_4 e^{1.3 X_4} + \alpha_5 X_8 + \alpha_6 X_9 + \alpha_7 X_{10}$$

- **Outcome 3.**

$$Y = \alpha_0 + \gamma T + 4 \sin(\alpha_1 X_1 + \alpha_2 X_2 + \alpha_3 X_3 + \alpha_4 X_4 + \alpha_5 X_8 + \alpha_6 X_9 + \alpha_7 X_{10})$$

- **Outcome 4.**

$$Y = \alpha_0 + \gamma T + \alpha_1 X_1 + \alpha_2 X_2^2 + \alpha_3 X_3 + \alpha_4 e^{1.3 X_4} + \alpha_5 X_8 + \alpha_6 X_9 + \alpha_7 X_{10}$$

The true values of the coefficients used are:

$$\gamma = -0.4,$$
$$(\alpha_0, \alpha_1, \alpha_2, \alpha_3, \alpha_4, \alpha_5, \alpha_6, \alpha_7) = (-3.85, 0.3, -0.36, -0.73, -0.2, 0.71, -0.19, 0.26).$$

*Outcomes 2, 3 and 4* correspond to outcome models 4, 5 and 2 of Setodji (Setodji et al. 2017), respectively.

*Appendix* A briefly presents the data generation steps, as well as the steps followed to obtain the values of balance evaluation reported below.

For the continuous outcomes, the outcomes 2 and 4 show a greater deviation from additivity compared to outcome 3. Thus, it is likely to be more difficult to obtain a good estimation of the causal treatment effect when using a simple linear outcome model that only controls for the pre-treatment confounders using main effects.

## Sets of Confounders

There is some discussion in the literature about the covariates one should consider as confounders and use to estimate the PS and balancing weights (see *section* 4). In order to test these suggestions, we consider several sets of confounders, each including different baseline covariates. These are:

- **Confounders-set 1: true_confounders.** This is the set of true only $(X_1, X_2, X_3, X_4)$ — covariates related both to the treatment status and the outcome.

- **Confounders-set 2: treatment_all.** This is the set of all covariates that are related to the treatment allocation $(X_1, X_2, X_3, X_4, X_5, X_6, X_7)$.

- **Confounders-set 3: all_covariates.** This set includes all covariates, either related to the treatment or the outcome $(X_1, X_2, X_3, X_4, X_5, X_6, X_7, X_8, X_9, X_{10})$ — all the available covariates.

- **Confounders-set 4: outcome_all.** This is the set of covariates that are related to the outcome $(X_1, X_2, X_3, X_4, X_8, X_9, X_{10})$.

- **Confounders-set 5: true_subset.** This set includes only two, out of four in total, true confounders $(X_1, X_2)$ — a restricted set of confounders.

- **Confounders-set 6: treatment_only.** This set includes the three covariates related to the treatment allocation only $(X_5, X_6, X_7)$ — no covariate related with the outcome is included.

- **Confounders-set 7: outcome_only.** This set includes the three covariates related only to the outcome only $(X_8, X_9, X_{10})$— no covariate related with the treatment allocation is included.

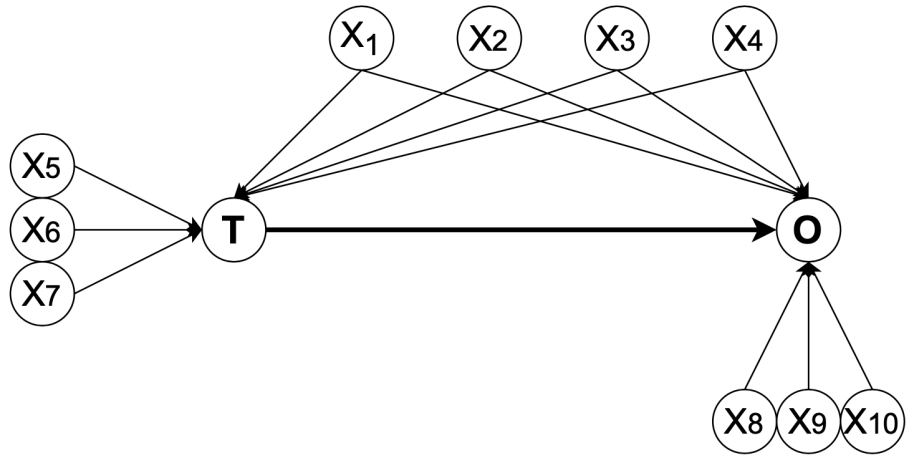

*Figure 1: Diagram representing the relation of each covariate with treatment/outcome. The arrows represent which baseline covariates affect the treatment/outcome.*

*Figure* 1 depicts the original relationship between the baseline covariates and the treatment/outcome covariates (main plot, top-left), while the subplots represents the relations each confounders-set considers — only the solid-arrow covariates are considered in each case.

These seven *confounders-sets* represent all the realistic potential combinations of covariates that could be included in the estimation of PS and balancing weights — covariates related exclusively with the outcome (confounders-set: *outcome_only*), related exclusively with the treatment allocation (confounders-set: *treatment_only*), the true confounders (confounders-set: *true_confounders*), all the covariates related to the treatment allocation (confounders-set: *treatment_all*), all the covariates related to the outcome (confounders-set: *outcome_all*), covariates related either with the treatment allocation or the outcome (confounders-set: *all_covariates*) and a part of the true confounders (confounders-set: *true_subset*).

## Sample Size

It is of key interest for us to understand the type of sample sizes required to draw good inferences in the space of PS and balancing weights. No rules of thumb have been developed that help guide researchers as to how many units/individuals is sufficient. The rule of $10-to-1$ (10 units/individuals per confounder per group) commonly used for regression models may not be sufficient in this situation and we therefore designed our simulations to allow for a more careful assessment of the impact of sample size on performance of PS weighting methods. Specifically, we evaluate the performance of the algorithms on sample sizes equal to $n = 40, 80, 100, 200, 300, 400, 500, 600, 800, 1000, 1500, 2000$, for every set of confounders, for every outcome.

# 6. Results

We estimate PS and balancing weights for each dataset (for the 1000 replicates of each simulated scenario), and estimate the ATT using *doubly robust estimator* (Bang and Robins 2005; Kang, Schafer, and others 2007; Chattopadhyay, Hase, and Zubizarreta 2020; Zhao and Percival 2016) — this is the PS and balancing weights are used as weights in an augmented regression model (logistic for binary and linear for continues outcomes, respectively). Outcomes 1 and 3 (see *section* 5.2) are very close to typical logistic and linear regression, respectively, thus it is possible to obtain a good estimator of the causal treatment effect using simple regression models without PS and balancing weights — here we consider a good estimator to be an estimator with low absolute relative bias. In such cases, an unweighted regression may be preferable as the ESS is not reduced, thus providing the maximum possible power. Of the remaining two outcomes (outcomes 2 and 4), outcome 2 shows the greater deviation from linearity. When analysing real data, the true relation between the baseline covariates and the outcome is not known, and cannot be ascertained. We therefore report results obtained from outcome 2, as it represents a situation where there is extreme mis-specification of the outcome model, and thus a "worst-case" scenario. Results concerning outcome 4 lead to similar conclusions. We do not discuss results of outcome 1, 3 and 4 in the main body of this article, however, we provide the corresponding of *figure* 4 in *Appendix* B — conclusion concerning outcome 4 are identical to the findings about outcome 2.

We begin the presentation of the results by comparing the relative performance of balance measures (*section* 6.1) — mean SMD, maximum SMD, mean KS and maximum KS — in predicting the absolute relative bias. We use these results to obtain the optimal balance statistic and threshold (*section* 6.2) which provides the most evidence that the causal treatment effect estimate will yield the lowest possible bias. Then, we discuss the performance of different sets of confounders with respect to absolute relative bias (*section* 6.3) — for a range of sample sizes —, and conclude with the number of units/individuals per covariate per group required (*section* 6.4) to obtain a treatment effect estimation with high accuracy.

## The Proper Balance Statistic

To understand which balance measure is most suitable for providing evidence of the balance achieved among the baseline covariates for the two groups, we examine the relationship between our four balance statistics (the mean SMD, maximum SMD, mean KS statistic, and maximum KS statistic) and the absolute relative bias of the estimated causal treatment effect. For the results reported on this section, we used confounders-set: *all_covariates* — all the available baseline covariates.

*Figure* 2 depicts mean SMD (bottom-right), maximum SMD (top-right), mean KS (bottom-left), maximum KS (top-left) on the $x-axis$, and the absolute relative bias on the $y-axis$.

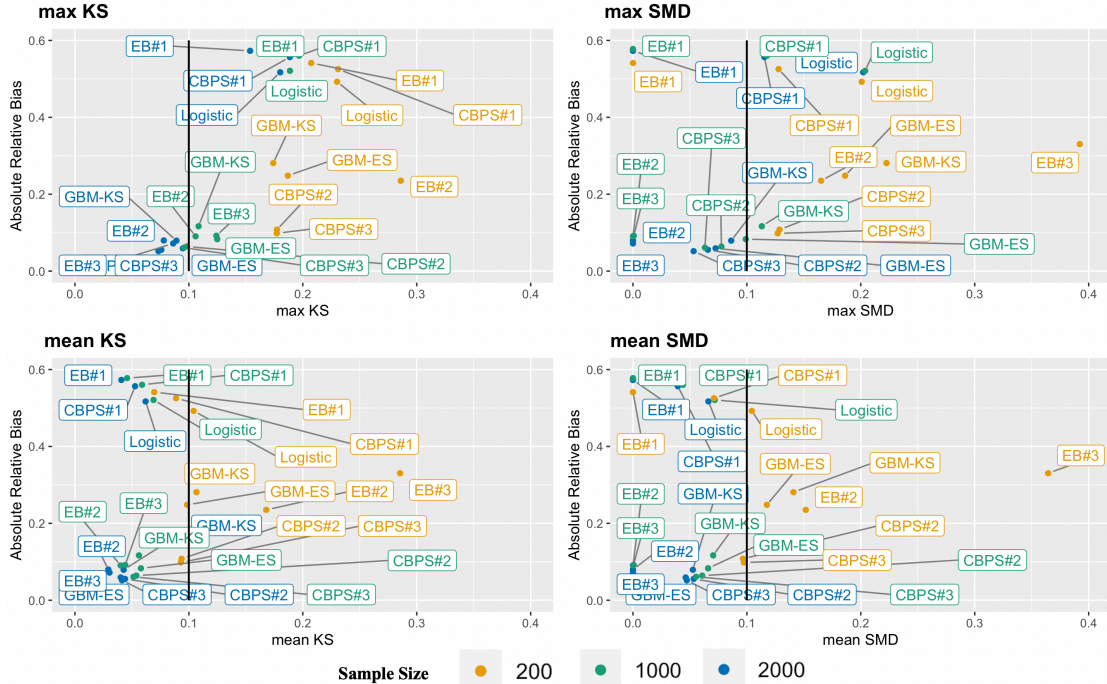

*Figure 2: Maximum KS statistic (top-left), maximum SMD (top-right), mean KS statistic (bottom-left), mean SMD (bottom-right) vs absolute relative bias of the estimation of the causal treatment effect. Each color represents a different sample size (*200,1000,2000*). The vertical black solid line represents the* 0.1 *level of balance, which is widely used as an example of acceptable balance. Each point corresponds to a different PS and balancing weights algorithm (labeled).*

From the figures concerning both maximum and mean SMD (right side), it is apparent that some algorithms which achieve exact balance (exactly zero mean difference) fail to estimate the causal treatment effect with a reasonable absolute relative bias — reported value over 0.5. Thus, low SMD is not a guarantee of low absolute relative bias. For example, Entropy Balance controlling for the first moment (*EB*#1), does not achieve low absolute relative bias, despite being an algorithm designed to achieve exact balance between the treatment groups on the mean value of each confounder. This seems to be the case independently of the sample size (we tested the case for sample size equal to 20000 and 50000). The reason for this is that *EB*#1 produces weights that are extremely close to (or exactly) 0, to achieve balance on the first moment between the treatment groups. By assigning 0 weights to some individuals, these are also excluded from the estimation of the causal treatment effect, i.e. are considered as non-significant observations. Thus, excluding some individuals from the estimation of treatment effect could bias the estimate. The same pattern is also observed when considering the mean KS statistic (bottom-left).

Conversely, maximum KS statistic is a very conservative measure, as it will report no balance if only one covariate reports a high value, however, it is the only balance statistic that has a consistent relationship with absolute relative bias — the higher the maximum KS statistic, the higher the absolute relative bias. In particular, algorithms achieving maximum KS <0.1 also achieve absolute relative bias of < 0.1. This suggests that maximum KS of 0.1 is a sufficient criterion to guarantee low absolute relative bias.

## Balance Threshold

Next we will explore the validity of using 0.1 as a threshold for achieving balance for *max KS* across a range of sample sizes and weighting algorithms. In this section we again focus on the results produced based on using confounders-set: *all_covariates* — this is the case where all covariates are considered as confounders, and thus all available information is taken into account when we compute the PS and balancing weights. We utilise this *confounders-set*, since it is the one with the highest number of confounders, thus it would require a larger sample to achieve balance. *Figure* 3 depicts the relationship of maximum KS statistic and absolute relative bias to sample size (40 to 2000).

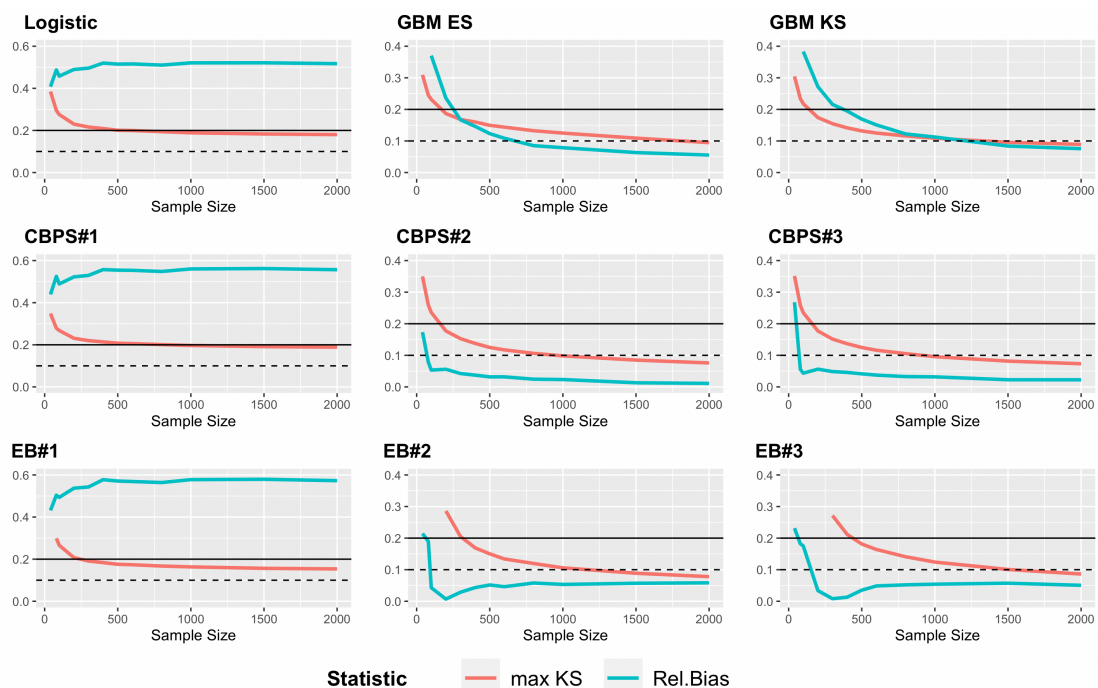

*Figure 3: Relationship of maximum KS statistic (red line) and absolute relative bias (blue line) to sample size (*40 *to* 2000*). Each sub-figure corresponds to a different algorithm for the estimation of the PS and balancing weights. The horizontal black solid line corresponds to a balance (or bias) threshold of* 0.2*, while the horizontal black dashed line corresponds to a threshold of* 0.1*.*

The horizontal solid black line corresponds to a balance threshold of 0.2 (Abdia et al. 2017), and it is apparent that PS and balancing weights estimated with every algorithm manage to balance the baseline covariates on this level, given a sufficient sample size. However, only PS and balancing weight algorithms which eventually reduce the maximum KS statistic below 0.1 (Griffin et al. 2017; Li and Li 2021; Setodji et al. 2017) (horizontal black dashed line) control the relative bias (i.e. $< 0.1$).

LR, *CBPS*#1 and *EB*#1, never produce PS and balancing weights with a maximum KS statistic less than 0.1, and the corresponding absolute relative bias is large for all sample sizes — this also true for sample sizes $> 2000$. These algorithms use parametric modeling to fit a model on first-order covariates. As a consequence, it is difficult to capture the higher-order relationships between the baseline covariates and the treatment allocation, which exist in the true propensity score model (see *section* 5.1). $CBPS\#2$, $CBPS\#3$, $EB\#2$ and $EB\#3$ manage to produce a maximum KS statistic below 0.1 for sample sizes greater than 1000, and low absolute relative bias ($< 0.1$) at lower sample sizes. This set of algorithms use parametric models with restrictions imposed to all $m$ first moments of the covariates (in our case $m = 2,3$). This helps these algorithms to fit the true underlying relationship between the baseline covariates and the treatment allocation. $GBM_{ES}$ and $GBM_{KS}$ both require a larger sample (about 2000) to achieve a low maximum KS statistic and absolute relative bias (below 0.1), which is expected considering the nature of the GBM algorithm. Since it is fitting trees to predict the probability of allocation to the treatment group, and the total number of iterations is typically high, a large sample is required.

## Variable Selection

*Figure* 4 shows the relationship between absolute relative bias of the causal treatment effect estimate and sample size (ranging from $n = 40$ to $n = 2000$), for each confounders-set separately. Each line represents a different algorithm. It is apparent that when the true confounders $(X_1, X_2, X_3, X_4)$ are not included in the PS and balancing weights model (and the outcome model) — confounders-set: *true_subset*, *treatment_only*, and *outcome_only*, then the estimation of the causal treatment effect always shows large absolute relative bias for all sample sizes and algorithms. High levels of bias in the causal treatment effect are also observed when the algorithm used for the estimation of the causal treatment effect does not achieve adequate balance on the baseline covariates, independently of the covariates included in the PS and balancing weights model. Indeed, $LR$, $CBPS\#1$, and $EB\#1$ do not balance the maximum KS statistic on the baseline characteristics (the value remains always above 0.1), and show a large absolute relative bias, independently of the covariates treated as confounders.

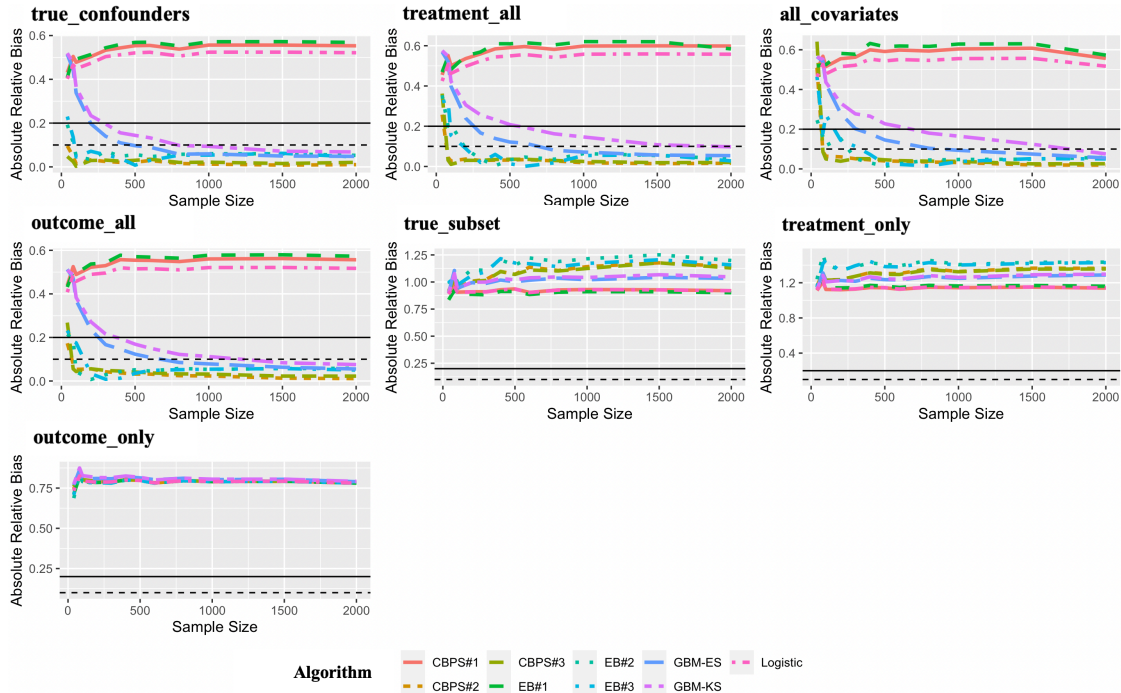

*Figure 4: Absolute relative bias ($y-axis$) vs sample size ($x-axis$), per confounders-set (see 5.3). Each colored line represents a different algorithm used for the estimation of the PS and balancing weights. The horizontal black solid line represents* 0.2 *level, while the horizontal black dashed line is* 0.1 *level.*

*Figures* 5 and 6 show the relationship between sample size (ranging from $n=40$ to $n=2000$) and the absolute relative bias and precision (the (MSE)), respectively, of the causal treatment effect estimate. These calculations were restricted to algorithms which achieve balance as assessed by the maximum KS statistic ($CBPS\#2$, $CBPS\#3$, $EB\#2$, $EB\#3$, $GBM_{ES}$ and $GBM_{KS}$). Each sub-figure shows one of the confounders-sets which include the — confounders-sets: *true_confounders*, *treatment_all*, *all_covariates*, *outcome_all* (see 5.3).

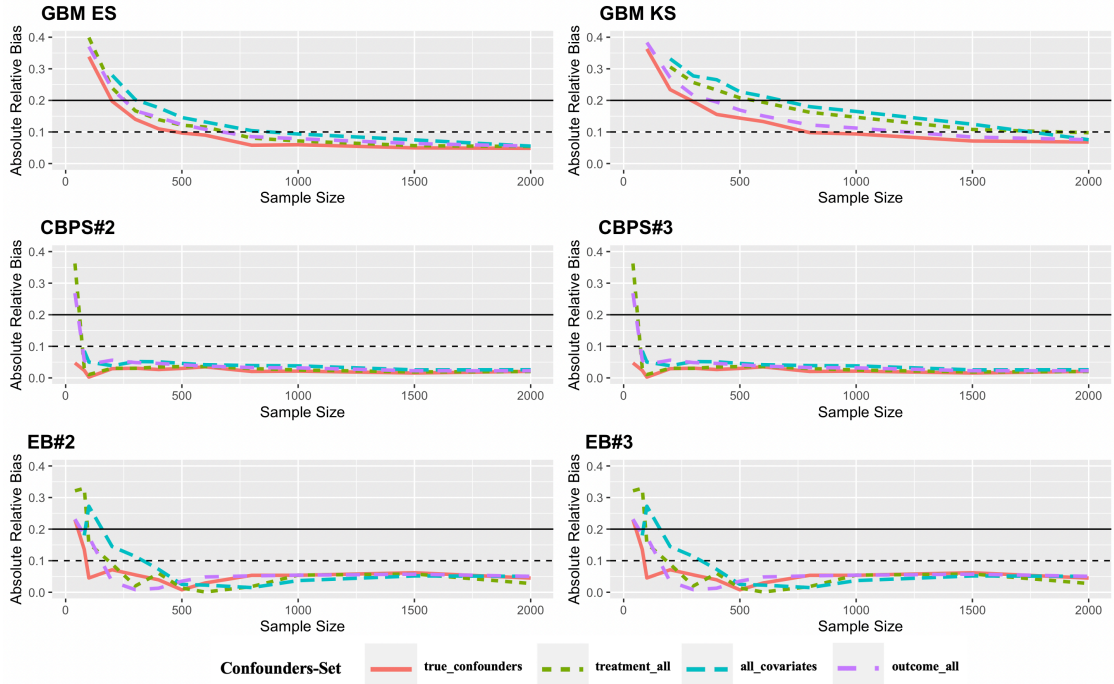

*Figure 5: Absolute relative bias (y − axis) vs sample size (x − axis), per algorithm (see 5.3). Each colored line represents a different confounders-set (confounders-set: true_confounders, treatment_all, all_covariates, outcome_all). The horizontal black solid line represents* 0.2 *level, while the horizontal black dashed line is* 0.1 *level.*

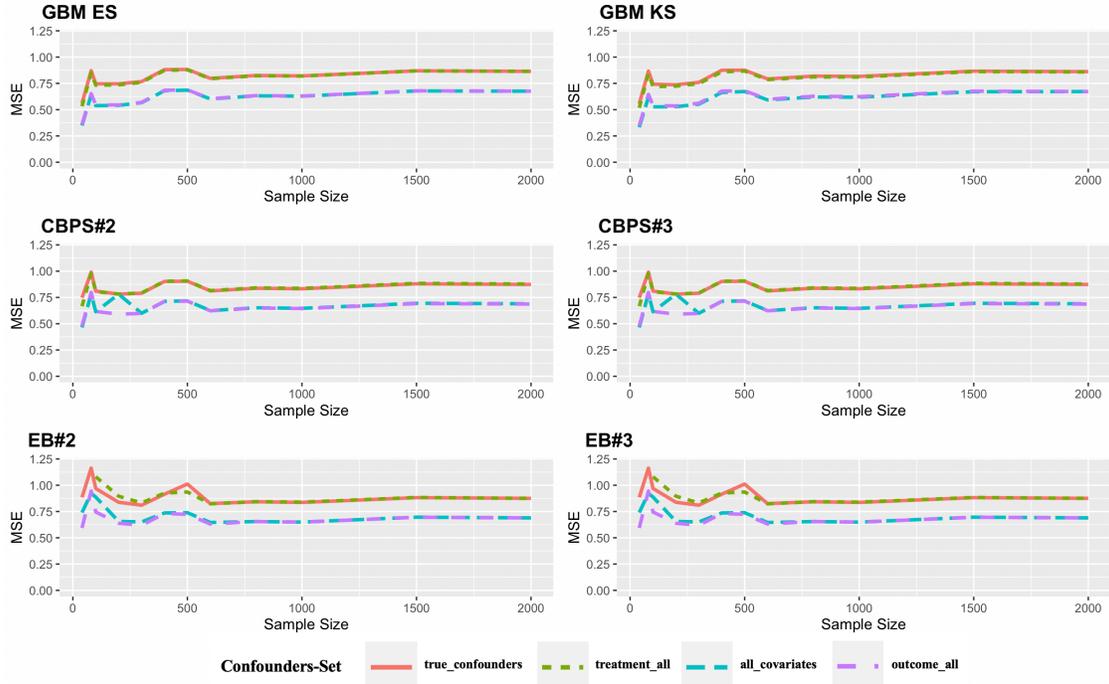

*Figure 6: Mean Square Error (y − axis) vs sample size (x − axis), per algorithm (see 5.3). Each colored line represents a different confounders-set (confounders-set: true_confounders, treatment_all, all_covariates, outcome_all).*

Although all confounders-sets that include the $(X_1, X_2, X_3, X_4)$ show similarly low absolute relative bias, it can be seen that confounders-sets *true_confounders* and *outcome_all* show the lowest values of absolute relative bias, particularly for small sample sizes. Additionally, it is apparent that the sets of confounders that include the covariates related only to the outcome (confounders-sets *all_covariates* and *all_covariates*) report lower MSE — thus, such models have better prediction ability. Inclusion of covariates related only to the outcome (confounders-set: *outcome_all*) contributes more in lowering bias than the inclusion of covariates related only to the treatment allocation (confounders-set: *treatment_all*), when added to the true confounders, which corroborates results in the literature (Brookhart et al. 2006; Patrick et al. 2011). Adding covariates which are related only to the treatment allocation and not to the outcome to the set of covariates considered for confounding bias (confounders-set: *treatment_all*), inflates the bias and the MSE, and also makes it harder for any algorithm to balance the baseline covariates since the number of covariates increases (Adelson et al. 2017) — compared to confounders-set: *true_subset*. Overall, confounders-sets *all_covariates* (all the available covariates) and *outcome_all* (the true confounders and covariates related only with the outcome) achieve the best trade-off between low bias and low MSE, thus — if information regarding the relation of the baseline covariates to treatment/outcome is available, and sample size is sufficient — these are the covariates one should use to control for confounding bias.

### Sample Size

*Figures* 7 and 8 show the relationship between maximum KS and absolute relative bias, respectively, with the sample size per confounder per group $(x - axis)$. We focus on confounders-sets *true_confounders*, *treatment_all*, *all_covariates* and *outcome_all* since when the true confounders are not included on the set of covariates one controls for confounding bias, the estimation of the causal treatment effect reports extremely high absolute relative bias — each line represents a different confounders-set (see *section* 5.3). We consider *CBPS*#2, *CBPS*#3, *EB*#2, *EB*#3, $GBM_{ES}$ and $GBM_{KS}$, as these algorithms achieve an adequate balance (maximum KS statistic below 0.1), and low absolute relative bias.

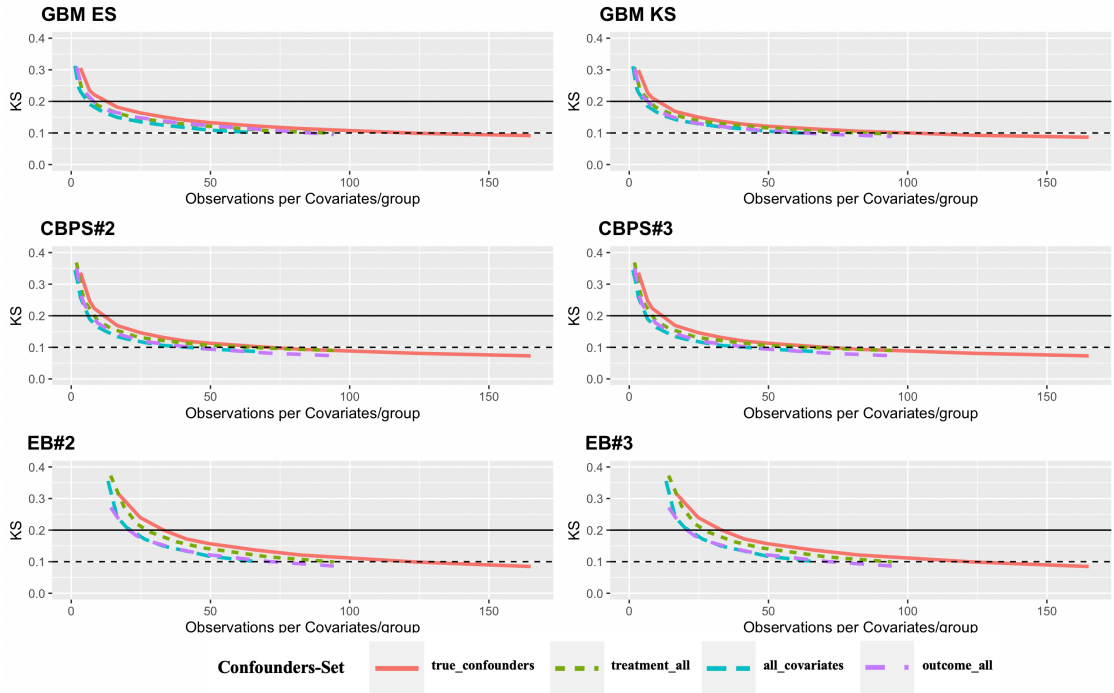

*Figure 7: Sample size per confounder per set, vs maximum KS statistic. Each line represents a different confounders-set (see section 5.3). The horizontal black solid line represents* 0.2 *level, while the horizontal black dashed line is* 0.1 *level.*

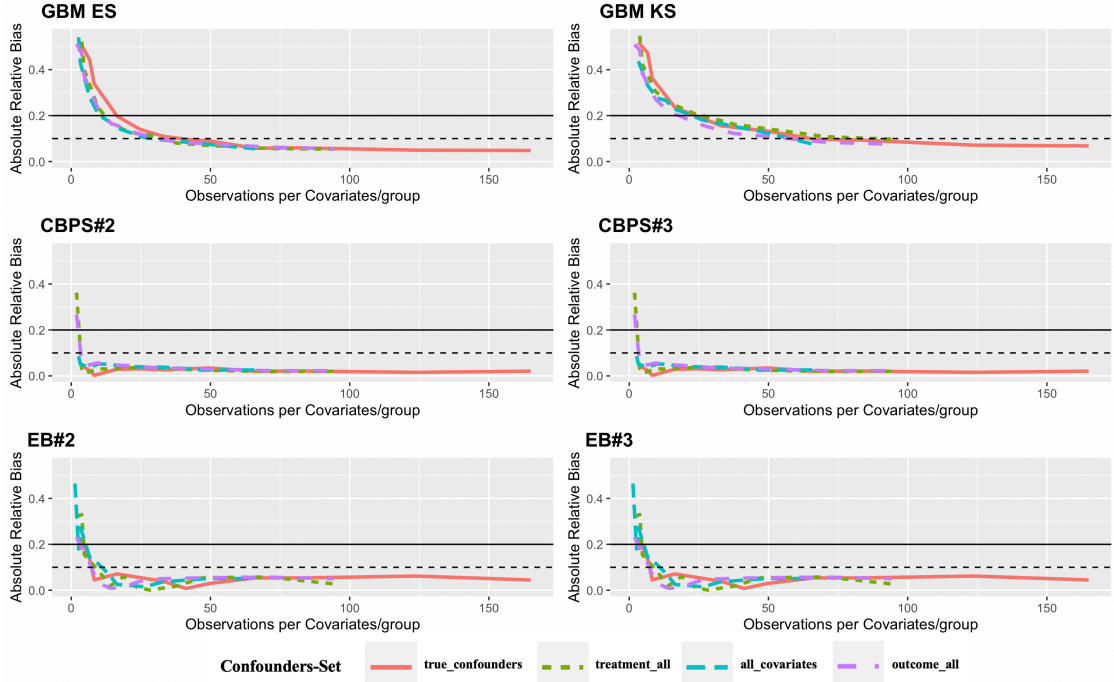

*Figure 8: Sample size per confounder per set, vs absolute relative bias. Each line represents a different confounders-set (see section 5.3). The horizontal black solid line represents* 0.2 *level, while the horizontal black dashed line is* 0.1 *level.*

The required number of observations per covariate per group to achieve adequate balance (< 0.1), as assessed by the maximum KS statistic (*figure* 7), is similar for all algorithms. Approximately $60 - 80$ observations per covariate per group are required to achieve a maximum KS statistic value below the threshold of 0.1, with the upper limit of this range required when the minimum number of covariates (the *true confounders* only — confounders-set: *true_subset*) is included in the PS and balancing weights model. Slightly fewer observations per covariate per group are required when more covariates are incorrectly assumed to be true confounders — however, this increases the overall sample size.

*Figure* 8 shows a similar pattern, when PS and balancing weights are estimated by machine learning algorithms ($GBM_{ES}$ and $GBM_{KS}$). Parametric algorithms ($CBPS\#2$, $CBPS\#3$, $EB\#2$ and $EB\#3$) require slightly lower numbers ($45 - 60$) of observations per covariate per group to achieve low bias (<0.1) of the causal treatment effect estimate. However, this advantage of the parametric model could be due to the similarity of the true PS model to the parametric models these algorithms fit. When both the true PS model and the true relationship between the baseline covariates and the outcome are mis-specified — an assumption that we are unable to check when dealing with real data —, the balance statistic is the only measure we have to assess the similarity of the groups on the baseline characteristics, and thus to make inference with higher confidence. In real data problems, it is not feasible to check whether the true relationship between the treatment status and the covariates we use to control for confounding bias is close to the relationship assumed on any parametric model, no matter how many higher-order moments or interactions one could include in the model. It is therefore not feasible to know whether the bias of the obtained estimation of the causal treatment effect is low unless sufficient balance has been achieved. In view of this, we recommend the use of several algorithms to compute PS and balancing weight (Markoulidakis et al. 2021) — using both parametric and non-parametric algorithms — and use for outcome analysis the weights that achieve the best trade-off of lower maximum KS statistic and larger ESS (Markoulidakis et al. 2021).

## 7. Discussion/Conclusion

Observational studies are becoming more and more popular given the greater use of routine data in applied health research. As such, PS and balancing weights are an area of active investigation. The estimation of the causal treatment effect using PS and balancing weights is a robust method to make inference based on data from observational studies, however, there are certain criteria that should be met to guarantee an unbiased and robust estimation of the desired estimand.

The choice of algorithm to compute PS and balancing weights is an important part of the analysis. Since the true relationship between the baseline covariates and the treatment status is usually unknown, we strongly recommend using multiple algorithms for the estimation of the PS and balancing weights (Markoulidakis et al. 2021). Of the algorithms that reduce the maximum KS statistic below 0.1, the one with with the highest ESS should be chosen, to maximise power. A potential advantage of GBM over the other algorithms is that it makes no parametric model assumptions for the relationship for the treatment allocation mechanism. As such, it is capable of estimating PS and balancing weights for more complex treatment allocation models.

In order to control for confounding bias, it is important to achieve a sufficient level of balance among the baseline covariates. The threshold of maximum KS statistic less than 0.1 seems to be adequate to guarantee low bias of the estimate of treatment effect, in contrast to 0.2, which is insufficient. Additionally, special care should be taken when selecting variables to control for confounding bias. If the selected covariates are not predictors of the outcome (i.e. confounders-set: *treatment_only*), the estimation of the outcome could have a huge bias, even if seemingly adequate balance is achieved. The best trade-off between bias and the precision of the estimate (MSE) is achieved when covariates that are related both to the treatment allocation and the outcome (the true confounders), and the covariates related only to the outcome, are included in the PS and balancing weights model (i.e. confounders-set: *outcome_all*), to control for confounding bias (Brookhart et al. 2006; Hirano and Imbens 2001; Perkins et al. 2000). Inclusion of the covariates related only to the treatment could improve the balance on the baseline covariates but could increase the bias (Brookhart et al. 2006; Hirano and Imbens 2001; Perkins et al. 2000). Finally, large sample sizes are required to guarantee the robustness of the estimate. It seems that an empirical rule is to have $60 - 80$ individuals per treatment group, per covariate that one wishes to control for — i.e. sample size $840 - 1120$ when controlling for 7 covariates.

Results on outcome model 3 indicate that when the true outcome model is close to the regression model used for the estimation of the causal treatment effect, smaller samples are acceptable, and a more liberal level of balance is not a restriction to obtain a valid estimate. In this case, even an unweighted estimate would perform well because the regression model is a good fit. However, for real data we are not able to know the true underlying relationship, thus all balance criteria should be strictly met, to be confident about the validity of causal treatment effect estimates.

In real case data, the sample size is often a limiting factor on the number of covariates one could include in the PS and balancing weights model to control for confounding bias (see 6.4). Additionally, since it is not known *a priori* which baseline covariates are related only to the treatment allocation, related only to the outcome, and which are related to both (the *true confounders*), we recommend discussing with subject-matter experts and utilise their prior knowledge of the relationship between the baseline characteristics and the treatment/outcome. Correlation coefficients (preferably *Spearman's Correlation* (de Winter, Gosling, and Potter 2016; Akoglu 2018), as it makes fewer assumptions on the model relationships between the study covariates) can be calculated for each baseline covariate with the treatment covariates and outcome variable. A combination of data-based evidence about the relationship of baseline covariates and treatment/outcome, and prior expert knowledge should be used to decide which covariates to treat as confounders. Correlation coefficients could also reveal underlying relationships between baseline covariates (e.g. $X_2$ and $X_6$ are highly correlated with correlation coefficient 0.9), and thus could be used to reduce the number of confounders included in the PS and balancing weights model, resulting in better balance.

Direct acyclic graphs (DAGs) (Textor et al. 2016) are another tool that could be used, supplementary to the correlation matrix and scientific advice, to decide which covariates should be included in the PS and balancing weights model. Causal diagrams (Textor and Liskiewicz 2012) are graphs that depict the relation between confounders, treatment, outcome, instrumental variables, and more. DAGs (Textor et al. 2016; Textor and Liskiewicz 2012; der Zander, Liśkiewicz, and Textor 2015) are graphical tools that work on causal diagrams, and could propose equivalent minimal sufficient diagrams, in the sense that any potential unnecessary paths could be removed. However, valid and adequate interpretation of these graphic tools is vital, and any decision about the final set of confounders should be communicated with scientific advisors of the study and clinicians, to guarantee that any modification will not affect the estimation (and the interpretation) of the causal treatment effect.

Previous work (Hirano and Imbens 2001) suggests that goodness-of-fit could be deployed to select which covariates to control for confounding bias — this is the typical backward selection of covariates on regression models, starting from a full set of covariates, gradually removing one covariate at a time until only statistically significant covariates remain in the final set. The caveat of such a strategy is that one can only identify which covariates are related to the treatment and which are related to the outcome (separately), and this holds only as long as the regression model used for the significance test is close to the true relationship between the response and the regressors. As a consequence, the methods referred to above for variable selection seem more advisable, since they take into consideration the scientific knowledge and the inherent effect of the confounders on both the treatment allocation and the outcome.

In order to guarantee good estimation of the causal treatment effect, one needs a large enough sample, a proper set of covariates to control for confounding bias (the true confounders and the covariates related only to the outcome), and a strict threshold to evaluate balance (maximum KS less than 0.1 is recommended). If any of these three conditions is not met, then it is possible to obtain an estimate with large bias and MSE — indicating that this estimation is not robust to unobserved confounders.

## Funding Support

DOMINO-HD (the project that funds the first author's PhD) is funded though the EU Joint Program for Neurodegenerative Disease Research with UK funding from Alzheimer's Society and Jacques and Gloria Gossweiler Foundation. The Centre for Trials Research, Cardiff University receives infrastructure funding from Health and Care Research Wales. This work was also supported by Medical Research Council (UK) grant MR/L010305/1. Funding was also provided by grant R01DA045049 (PI Griffin) through the National Institute of Drug Abuse.

# Appendix A. Data Generation Process

1. First, six covariates, $(X_1, X_2, X_3, X_4, X_7, X_{10})$ were generated as independent standard normal random variables with zero mean and unit variance.

2. Then, another four covariates $(X_5, X_6, X_8, X_9)$ were generated with the described correlation to the other covariates (see above for the correlation matrix). These values refer to the magnitude of the correlation coefficient before dichotomizing some of the covariates $(X_1, X_3, X_5, X_6, X_8, X_9)$. Dichotomizing these covariates would attenuate these values. To dichotomize the covariates, first, a random number between 0 and 1 from a uniform distribution was generated. $X_i$ was set to be 1 if the randomly generated number was less than $1/2$, and as 0 if the number was greater than the estimated true propensity score.

3. The dichotomous exposure, $T$ was modeled using ([eq:treatment_eq]). First, a random number between 0 and 1 from a uniform distribution was generated. $T$ was set to be 1 if the randomly generated number was less than the estimated true propensity score $(P(T|X_i))$, and as 0 if the number was greater than the estimated true propensity score.

4. The outcome, $Y$, was modeled using ([eq:output_model_1]–[eq:output_model_4]). For the case of ([eq:output_model_1]), where the outcome is binary, a random number between 0 and 1 from a uniform distribution was generated. $Y$ was set to be 1 if the randomly generated number was less than the probability of $Y$ given $T$ and $X_i$ $(P(Y|X_i, T))$, and as 0 if the number was greater than the probability of $Y$ given $T$ and $X_i$.

5. In the step of the outcome simulation, we inserted the treatment effect, to be equal to $-0.4$ for individuals in the treatment group.

6. To obtain robust inference, we repeated the above procedure 1000 times, resulting in 1000 datasets for each set-up.

# Appendix B. Results: Outcome 1, 3 & 4

*Figure* 9, depict the evolution of maximum KS per confounders-set (see *section* 5.3), for each algorithm, as the sample size increases from 40 to 2000 observations. *Figures* 10, 11 and 12 depict the respective evolution of absolute relative bias as a function of the sample size, for outcome 1, 3 and 4, respectively (see *section* 5.2).

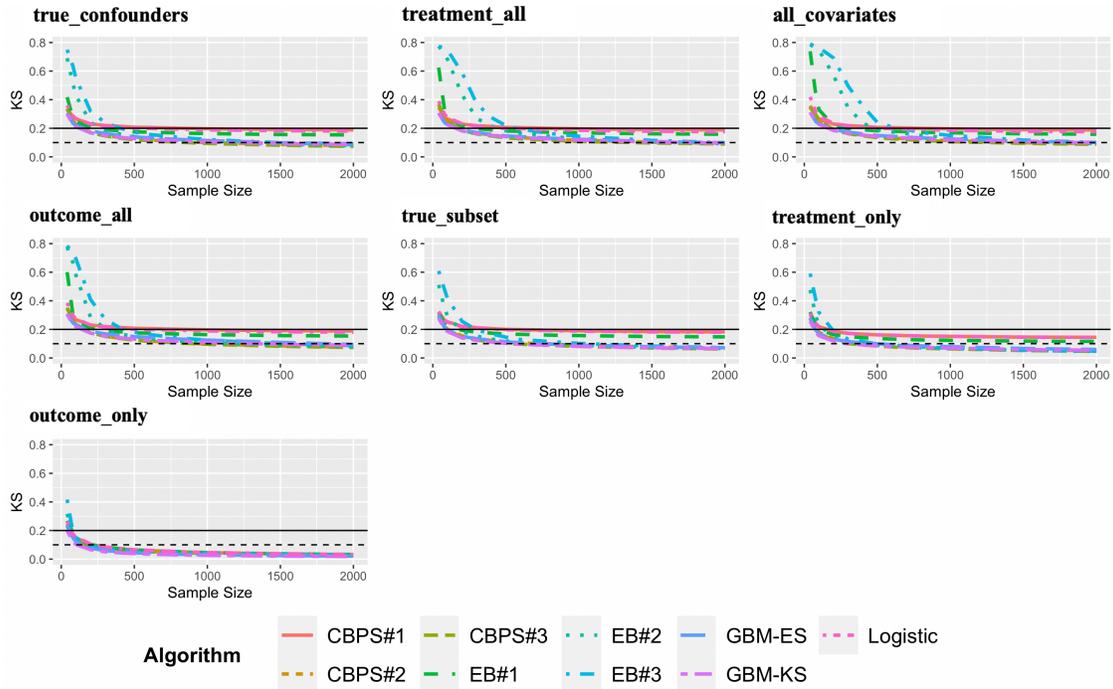

*Figure 9: Maximum KS (y − axis) vs sample size (x − axis), per confounders-set (see 5.3). Each colored line represents a different algorithm used for the estimation of the PS and balancing weights. The horizontal black solid line represents* 0.2 *level, while the horizontal black dashed line is* 0.1 *level.*

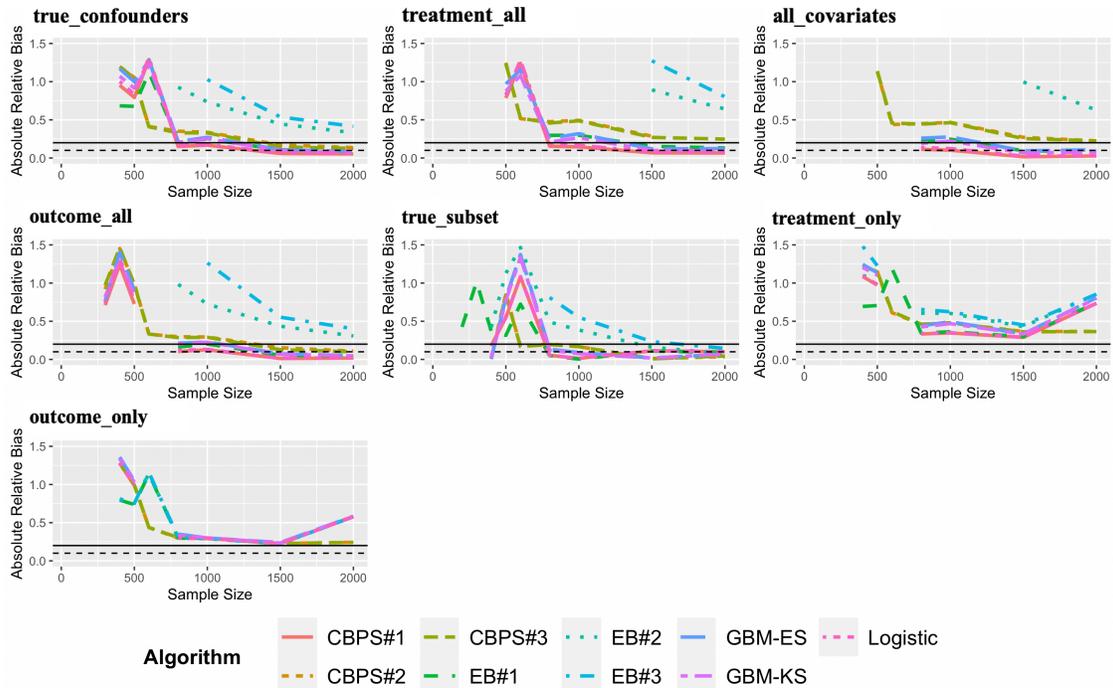

*Figure 10:* **Outcome 1:** *Absolute relative bias (y − axis) vs sample size (x − axis), per confounders-set (see 5.3). Each colored line represents a different algorithm used for the*

*estimation of the PS and balancing weights. The horizontal black solid line represents* 0.2 *level, while the horizontal black dashed line is* 0.1 *level.*

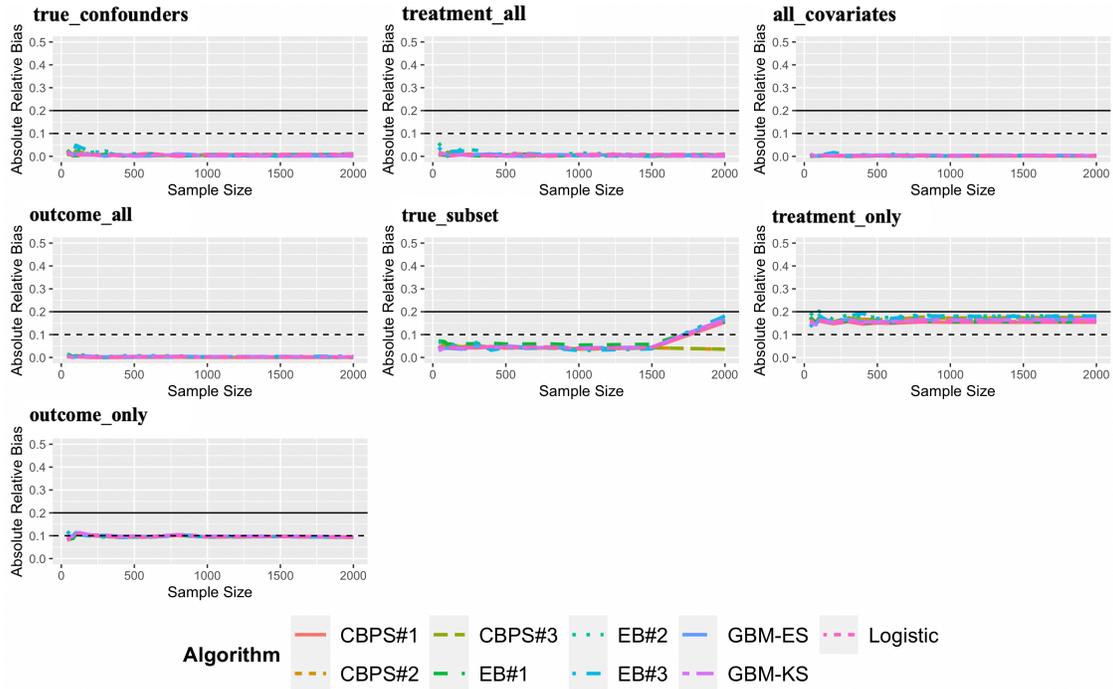

*Figure 11:* **Outcome 3:** *Absolute relative bias (y − axis) vs sample size (x − axis), per confounders-set (see 5.3). Each colored line represents a different algorithm used for the estimation of the PS and balancing weights. The horizontal black solid line represents* 0.2 *level, while the horizontal black dashed line is* 0.1 *level.*

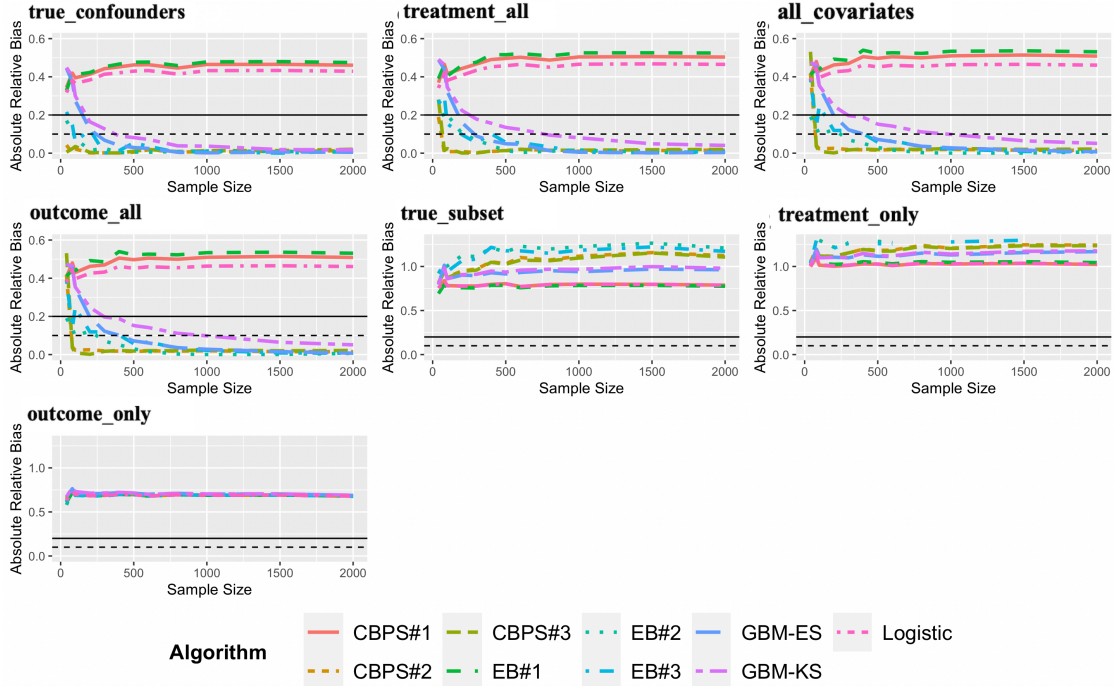

*Figure 11:* **Outcome 4:** *Absolute relative bias (y − axis) vs sample size (x − axis), per confounders-set (see 5.3). Each colored line represents a different algorithm used for the estimation of the PS and balancing weights. The horizontal black solid line represents* 0.2 *level, while the horizontal black dashed line is* 0.1 *level.*

From *figure* 11, which concerns outcome 3, it is apparent that every algorithm performs similarly in term of reported absolute relative bias. The value is typically on or below 0.1, no matter which algorithm is used to compute PS and balancing weights, or which confounders-set is used as predictors. The only exemption is when the covariates who are related only to the treatment and not related to the outcome are used as predictors (confounders-set: *treatment_only*). Even in this case though, the absolute relative bias is still below 0.2. the true relationship between outcome 3 and the baseline covariates (covariates $X_1, X_2, X_3, X_4, X_8, X_9, X_{10}$) is almost linear (see *section 5.2*), which means that the parametric (linear) model we use to estimate the causal treatment effect, is able to predict the treatment effect with high accuracy, independently of the balancing weight used and the balance level achieved by these weights between the treatment groups (see *figure* 9).

Estimation of causal treatment effect, using outcome 1, performs similarly (*figure* 10). In this case, since the outcome is binary (thus logistic regression is used in the outcome analysis to obtain an estimation of the causal treatment effect), a larger sample size is required to obtain an estimation with low balance — independently of the algorithm used to compute PS and balancing weights. However, again all algorithm perform similarly, no matter which confounders-set is used (see *section* 5.3). The relationship between outcome 1 and the baseline covariates could be perfectly modeled by the logit model, thus a good estimation of the causal treatment effect could be obtained even without the use of balancing weights, as soon as the model used for outcome analysis reflects the true relationship between the outcome value and the baseline covariates.

Conclusion about the performance the estimation of causal treatment effect, using outcome 4 (*figure* 12), are identical to those of outcome 2. Only the algorithms who manage to achieve balance — as this is evaluated from maximum KS statistic (balance threshold 0.1) — report an reasonable absolute relative bias. Again, only confounders-sets that include all the 4 true confounders (confounders-sets: *true_confounders*, *treatment_all*, *all_covariates* and *outcome_all*) make a good guess of the causal treatment effect — confounders-sets *true_confounders* and *outcome_all* report slightly lower absolute relative bias, compared to *treatment_all* and *all_covariates*.